\documentclass[journal=nalefd,manuscript=article]{achemso}  

\usepackage[version=4]{mhchem}
\usepackage[table,xcdraw]{xcolor}

\newcommand{\coo}{\ce{CO2} }

\title{Two-Dimensional Forms of Robust CO$_2$ Reduction Photocatalysts}

\author{Steven B. Torrisi}
\email{torrisi@g.harvard.edu}
\affiliation{Department of Physics, Harvard University, Cambridge MA, 02138}
\alsoaffiliation{Energy Technologies Area, Lawrence Berkeley National Laboratory, Berkeley CA 94720}
\author{Arunima K. Singh}
\affiliation{Department of Physics, Arizona State University, Tempe AZ 85287}
\alsoaffiliation{Chemical Sciences Division, Lawrence Berkeley National Laboratory, Berkeley CA 94720}
\author{Joseph H. Montoya}
\affiliation{Toyota Research Institute, Los Altos, CA 94022}
\alsoaffiliation{Energy Technologies Area, Lawrence Berkeley National Laboratory, Berkeley CA 94720}
\author{Tathagata Biswas}
\affiliation{Department of Physics, Arizona State University, Tempe AZ 85287}
\author{Kristin A. Persson}
\email{kapersson@lbl.gov}
\affiliation{Energy Technologies Area, Lawrence Berkeley National Laboratory, Berkeley CA 94720}
\alsoaffiliation{Department of Materials Science \& Engineering, University of California, Berkeley, Berkeley CA 94720}

\usepackage{lineno}

\begin{document}

\begin{abstract}
Novel photoelectrocatalysts that use sunlight to power the CO$_2$ reduction reaction will be crucial for carbon-neutral power  and energy-efficient industrial processes. Scalable photoelectrocatalysts must satisfy a stringent set of criteria, such as stability under operating conditions, product selectivity, and efficient light absorption. Two-dimensional materials can offer high specific surface area, tunability, and potential for heterostructuring, providing a fresh landscape of candidate catalysts. From a set of promising bulk CO$_2$ reduction photoelectrocatalysts, we screen for candidate monolayers of these materials, then study their catalytic feasibility and suitability. For stable monolayer candidates, we verify the presence of visible-light band gaps, check that band edges can support CO$_2$ reduction, determine exciton binding energies, and compute surface reactivity. We find for SiAs, ZnTe, and ZnSe monolayers, visible light absorption is possible, and reaction selectivity biases towards CO production. We thus identify SiAs, ZnTe, and ZnSe monolayers as targets for further investigation, expanding the chemical space for CO$_2$ photoreduction.

\end{abstract}

%%% INTRODUCTION TO CO2 CATALYSIS
%\section{Introduction}
Efficient, stable, scalable photoelectrocatalysts (PECs) which convert sunlight and \coo into useful products provide a desirable path towards achieving society's urgent carbon-neutral energy goals \cite{White2015,Chu2017a}.
Three example applications of \coo reduction products include (i) short-term storage of solar energy using methane \cite{Vogt2019a}, which forms a basis for decentralized solar electricity generation, (ii) generating syngas mixtures of CO and \ce{H2} as feedstocks for  the Fischer-Tropsch process\cite{Jahangiri2014}, or (iii) decreasing the carbon footprint of current industrial processes through efficient production of widely used feedstocks like formic acid. Efficient electrochemical reduction of \coo requires catalysts which can survive a strongly reducing environment and provide product selectivity at low overpotentials with respect to the complete reaction pathway \cite{Peterson2010}. Furthermore, photoelectrocatalysts must clear the same hurdles while also efficiently capturing light and providing photo-excited electrons at the appropriate \coo reduction energy. The search for \coo reduction PECs is thus an active and challenging area of research.

In this work, we present a comprehensive study of two-dimensional forms of recently suggested candidate \coo reduction catalysts. The chosen chemical systems were recently identified for their desirable properties in the bulk phase, such as stability under reducing conditions, suitable band structure, and appropriate band edges\cite{Singh2019}. Specifically, we target novel chemistries for \coo photoelectroreduction: the compounds ZnTe, ZnSe, GaTe, GaSe, AlSb, SiAs, YbTe, and AlAs form the basis of our investigation.
For these compounds, we explore the 2D structural landscape for new thermodynamically and dynamically stable monolayer structures,  and  evaluate their optoelectronic and reactive suitability as \coo reduction PECs.

 %CO2 PHOTOREDUCTION OVERVIEW
Photoelectrocatalysis is distinguished from photocatalysis and electrocatalysis by the mechanism of electron excitation and transfer. In photocatalysis, a molecule's potential energy surface is modified by adsorption onto the catalyst, allowing a photon to directly interact with the molecule and induce the desired reaction.  Electrocatalysis is characterized by the application of an electric potential $U$ on an adsorbing electrode, tuning the electron chemical potential and causing a desired reaction on the surface to become spontaneous and kinetically facile\cite{Nørskov2014}.  The focus of this study is on cathodic photoelectrocatalysis, in which an electron within the catalyst is excited by light which then transfers to an adsorbate and facilitates a reaction \cite{Maeda2007,Linic2015}. In other words, the energy source powering the reaction comes from the incident light instead of an applied bias voltage. 

The suite of desired properties for a photoelectrocatalyst -- the ability to absorb visible light, allowing reaction intermediates to bind, and generating photoexcited electrons of sufficient energy to trigger a reaction of interest -- fundamentally arise from the structure of a material  \cite{Hammer1995,Ulissi2016,Ulissi2017,Ulissi2017e,Tran2018}. Thus, unexplored structures may yield undiscovered functionality. Two-dimensional (2D) systems, characterized by a layered crystalline structure, offer a less well-studied and highly tunable avenue for future PECs \cite{Deng2016b,Singh2015}. The surprising effects of two-dimensional material interactions, such as ZnS/PbO and ZnSe/PbO heterostructures exhibiting enhanced solar absorption\cite{Zhou2018a}, expands the space of possible 2D-material based reactors even further beyond monolayers to the massive combinatoric space of heterostructures and 2D material coatings\cite{Hu2017,Deng2016b,Chae2017}. 

%INTRODUCE OUR WORK

\section{Results}
%% WORKFLOW SUMMARY
\subsection{Workflow}
\begin{figure}[htb!]
\includegraphics[width=13cm]{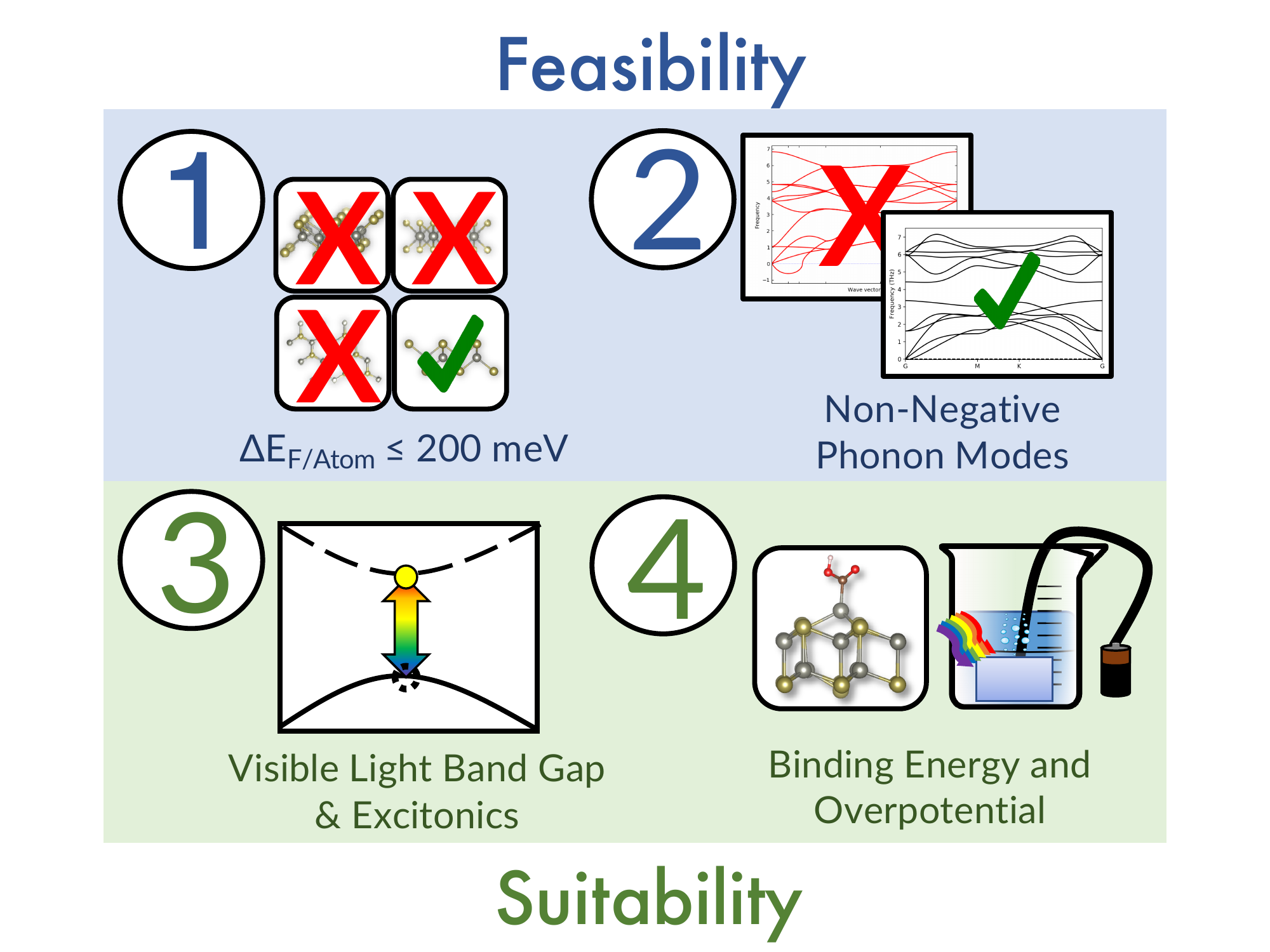}
\caption{Overview of workflow aimed at the following assessments for candidate structures: 1. Thermodynamic stability, as measured by comparing the formation energy of a monolayer against a cutoff. 2. Dynamic stability, as indicated by non-imaginary phonon frequencies in vacuum. 3. Solar photoreducibility, indicated by visible light band gaps and appropriate band edges, 4. Reactivity, as measured by the binding energy of adsorbates (COOH, CO), which are indicative of a \coo reduction pathway to form carbon monoxide.
\label{fig:screening}}
\end{figure}

To probe the diversity of 2D structure types for each compound and investigate the feasibility of \coo reduction, we devise a computational workflow (see Figure \ref{fig:screening}). Conceptually, the workflow addresses the \emph{feasibility} of candidate 2D structures of a given material, and the \emph{suitability} of these structures as PECs. We study feasibility by estimating the thermodynamic and dynamic stability of the structures, and suitability by assessing band gaps, band edges, exciton binding energies and adsorption energies of reaction intermediates.

%SUMMARIZE PREVIOUS WORK ON THESE MATERIALS
Previous experimental and theoretical works have investigated the structural, electronic, and chemical characteristics of a few monolayer phases of GaSe \cite{Tan2016}, GaTe\cite{Pozo-Zamudio2015}, ZnSe \cite{Sun2012a,Li2015e,Chaurasiya2019a,Zhou2018a,Lv2018,Zhou2017a,Zheng2015,Safari2017,Zhou2015b,Tong2014}, ZnTe \cite{Zheng2015,Safari2017,Tong2014}, and SiAs \cite{Cheng2018,Bai2018,Zhang2018c}. 2D layers of ZnSe in various structural forms have attracted attention for their photoabsorption properties \cite{Sun2012a,Zhou2017}, as have low-dimensional forms of ZnTe bound to nanowires or in nanoparticle form \cite{Jang2014,Jang2016,Zheng2018}. However, to our knowledge, none of the 2D systems we consider have been studied for \coo photoelectrocatalyst applications. 

%%%%%%%%%%%%%%%%%%%%%%%%%%%%%
% SCREENING FIGURE
%%%%%%%%%%%%%%%%%%%%%%%%%%%%%

\subsection{Feasibility Screening}
\subsubsection{Thermodynamic Screening}

% THE ORIGIN OF STRUCTURES 

%%%%%%%%%%%%%%%%%%%%%%%%%%%%%
% THERMODYNAMICS FIGURE
%%%%%%%%%%%%%%%%%%%%%%%%%%%%%

 \begin{figure}
 \includegraphics[width=12cm]{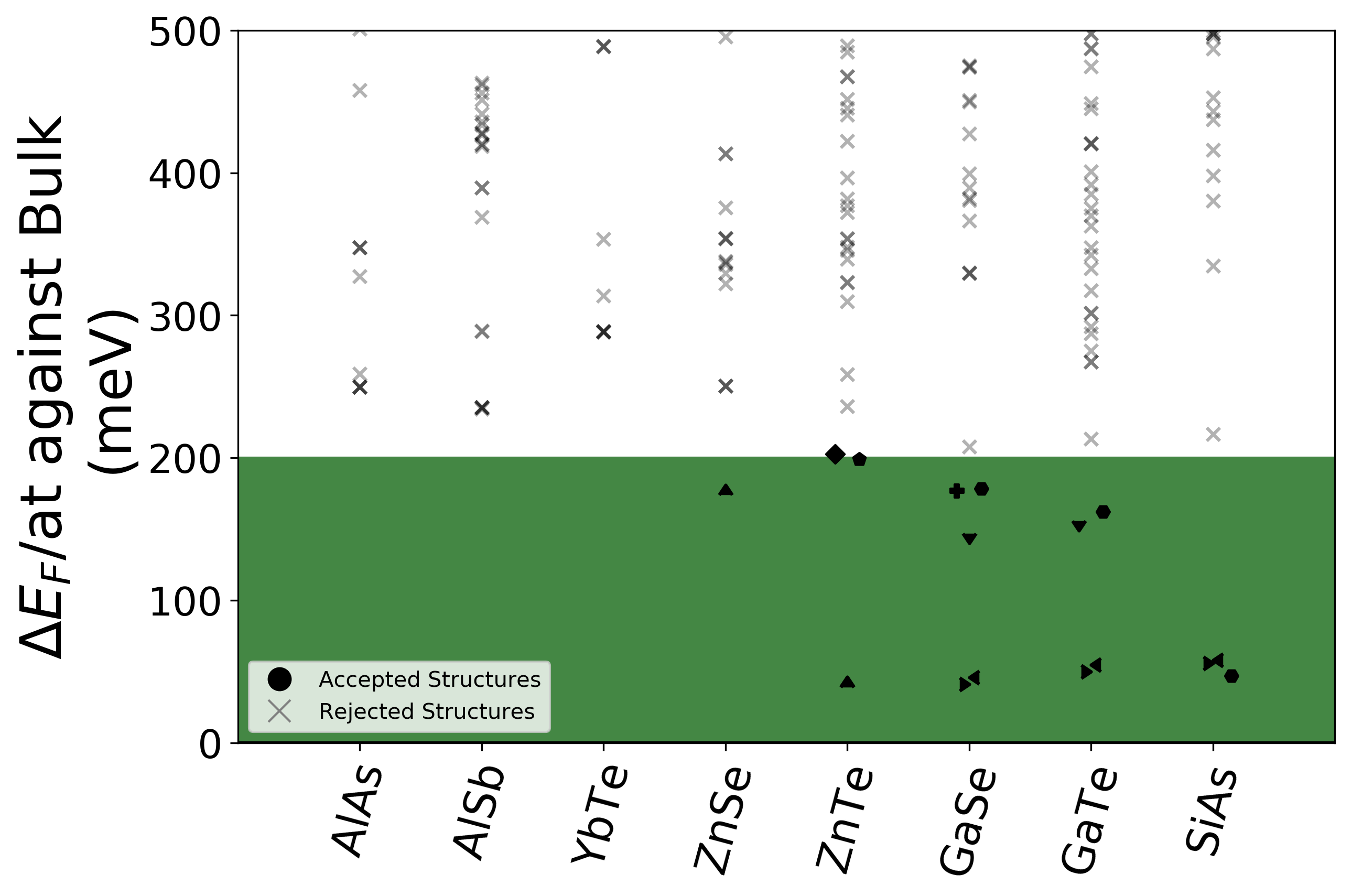}
\caption{Difference in formation energy per atom between a candidate 2D structure and the ground-state bulk, as computed using the SCAN+rVV10 functional. On the horizontal are bulk compounds we attempted to find 2D phases of. The horizontal offset of points within a column is for visual clarity.
On the vertical are differences in energy. Structures which are below our threshold are marked with symbols annotated in Figure \ref{fig:structures}, and those above it, with an X. \label{fig:energetics} 
}
\end{figure}

To ``seed'' our search for stable monolayers, we drew from the subset of 258 monolayer structures published by Mounet \latin{et al.} \cite{Mounet2018} as well as the 2D prototype structures from the Computational 2D Materials Database by Haastrup \latin{et al.} \cite{Haastrup2018}. We also included the native monolayer form of layered bulk SiAs. For our target materials, we ``mapped'' elements into monolayer structures of binary compounds (e.g. mapping ZnTe into the MoS$_2$ or h-BN structures). 

A recent data-driven study found the energy difference between the formation energy of the monolayer as compared to the equivalent bulk phase to be one of the most important predictors for MAX and MXene monolayer stability \cite{Frey2019}. Here, we define the change in formation energy, $\Delta E_F$,  as 
$$ \Delta E_\textrm{F} = E_\textrm{F,Monolayer} - E_\textrm{F,Bulk}$$
where $E_\textrm{F,Monolayer}$ is the formation energy of the 2D material
and $E_\textrm{F,Bulk}$ is the formation energy of its most stable 3D bulk counterpart. For e.g. for ZnTe, we used the structure mp-2176 in the Materials Project database \cite{Jain2013} to define the bulk formation energy, and normalize each formation energy per atom in the corresponding unit cell. Figure \ref{fig:energetics} shows the formation energies of the various 2D structures that were considered for the compositions AlAs, AlSb, YbTe, ZnSe, ZnTe, GaSe, GaTe, and SiAs. All formation energies were computed using density-functional theory with the SCAN+rVV10 functional. More details are available in the methods section. 

\begin{figure}
 \hspace{-1cm}
  \includegraphics[width=14cm]{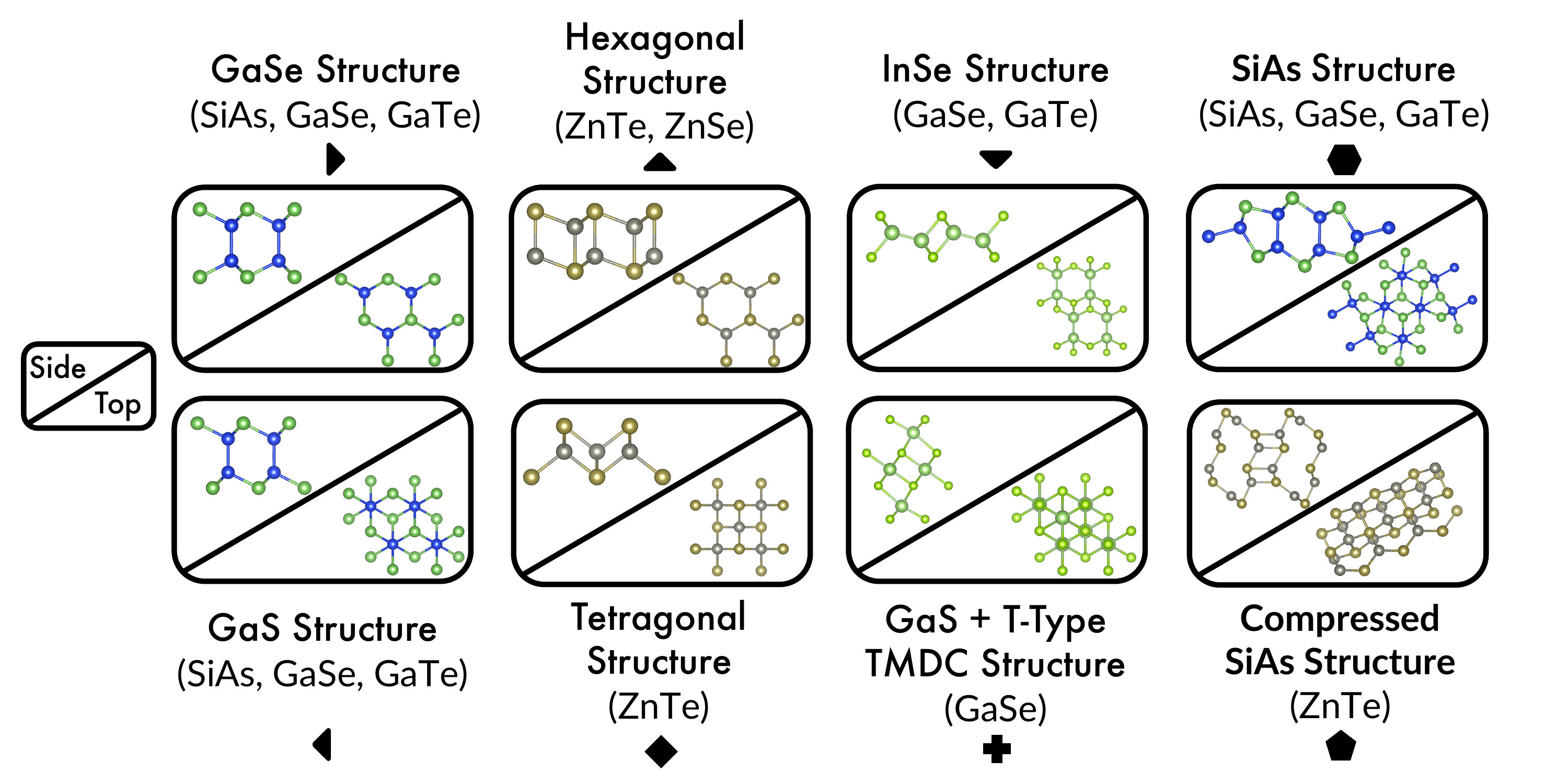}
  \caption{Visuals of structures which cleared the thermodynamic screening process depicted in Figure \ref{fig:screening}. Parentheses below each panel indicates the materials which met the thermodynamic criteria for the given structure, and a symbol which is used to refer to that structure in Figures \ref{fig:energetics} and \ref{fig:electronics}. Naming conventions for structures from \latin{Ashton et al.} \cite{Ashton2017}; we use the term ``compressed SiAs structure'' as it was the result of relaxation from the SiAs structure for ZnTe. Compounds pictured by column are: First column, SiAs (blue/green). Second column, ZnTe (silver/gold). Third column, GaSe (Ga large and green/Se small and light green). Fourth column, on top, SiAs, on bottom, ZnTe. \label{fig:structures}}
  \end{figure}

% ENERGETIC RESULTS
We use a a small $\Delta E_F \leq$ 200 meV/atom stability cutoff for our candidate monolayer structures, which Singh \latin{et al.} \cite{Singh2015} and Haastrup \latin{et al.} \cite{Haastrup2018} identified  as a useful heuristic stability criteria. The structures lying in the green shaded region of Figure \ref{fig:energetics} satisfy this stability criteria, and Figure \ref{fig:structures} depicts the stable structural prototypes. The $\Delta E_F \leq 200$ meV / atom criteria resulted in eight different structural prototypes for five of the compounds, and excludes three compounds: YbTe, AlSb, and AlAs. We found only one candidate structure for ZnSe, but three for ZnTe and SiAs, four for GaTe, and five for GaSe. The naming conventions for all the eight structure types are shown in Figure \ref{fig:structures}. Later in the study, we will focus on ZnTe and ZnSe in structural prototypes which resemble the CuI and the CuBr  structures\cite{Ashton2017}. For notational simplicity, we refer to them as the `hexagonal' and `tetragonal' structures.

%%%%%%%%%%%%%%%%%%%%%%%%%%%%%
% STRUCTURES FIGURE
%%%%%%%%%%%%%%%%%%%%%%%%%%%%%

Four  structural prototypes (the GaSe, GaS, InSe, and native SiAs structures) in Figure \ref{fig:structures}) feature chemical environments in which the anions are coordinated by three cations, and each cation by three anions and another cation. The bulk structures of GaSe, SiAs, and GaTe all feature exactly this coordination, explaining their compatibility with most of the monolayer structures with the same coordination. SiAs in the InSe structure, which features the same coordination pattern, comes close to the cutoff at 216 meV/atom above the bulk.

ZnTe and ZnSe in the bulk crystallize in the zincblende structure because they are `octet compounds' \cite{Mooser1986a} which attain chemical equilibrium by filling an eight-electron set of $s$ and $p$ valence orbitals. Additionally, the hexagonal and tetragonal prototypes can be understood respectively as a cleaved (110) plane and (100) plane of the bulk zincblende structure \cite{Zhou2015b}. Unlike the GaSe and GaS structures, the tetragonal and hexagonal structure types feature fourfold opposite species coordination, explaining their ability to support ZnTe. Note that monolayer ZnSe meets the thermodynamic cutoff only in one configuration, the hexagonal structure. Previous studies \cite{Zhou2015b,Tong2014,Li2015e,Lv2018} have found that ZnSe in the tetragonal structure is dynamically stable, though it did not meet our thermodynamic stability criterion and hence we did not include it in the next steps of our study; we found it to have $\Delta E_F =  250$ meV/atom above the bulk using SCAN+rVV10. It was found (See Table 1 of Zhou \emph{et al}. \cite{Zhou2015b}) using the PBE\cite{Perdew1996b} exchange-correlation functional that the tetragonal structure form had a higher energy per atom than the hexagonal form, which is consistent with our findings, but Li \emph{et al.} found using the HSE06 functional\cite{Heyd2003,Krukau2006} a lower energy for the tetragonal structure. \cite{Li2015e}

Low-dimensional GaSe \cite{AndresPenares2017,Rahaman2017,Pozo-Zamudio2015}, GaTe \cite{Pozo-Zamudio2015}, and ZnSe \cite{Sun2012a} have all been experimentally studied with structures similar to those which we found. We are not aware of any experimental evidence of two-dimensional SiAs or ZnTe, though both were featured in theoretical studies. \cite{Cheng2018,Bai2018,Zheng2015,Safari2017} However, previous computational efforts considered  different structural prototypes for ZnTe than in this work.

\subsubsection{Dynamical Stability Screening}

% WHY AND HOW WE DID THE PHONON CALCULATIONS
For all the structures which passed the thermodynamic screening, we sought to determine their dynamic stability. The full phonon band structures (See Methods section and supplemental material) evidenced dynamic stability in vacuum, with most structures yielding non-imaginary frequencies. In many cases, we noted small, imaginary acoustic phonon modes close to the zone center; however, in the majority of cases, these disappeared with increasing the size of the simulation cell. For ZnTe, ZnSe, and one GaTe structure, despite very large supercells (up to 8x8x1), we retained small instabilities in the elastic limit.
For these structures, perturbing along the negative displacement vector and then relaxing yielded in all cases a return to the starting structure, suggesting that the unstable elastic modes are artifacts of an insufficient supercell size and the known difficulty of fitting the quadratic $\Gamma$ z-direction acoustic phonon\cite{Mounet2018}. GaSe in the GaS + T-Type structure type evidenced broad imaginary phonon modes off the Brillouin Zone center, indicating a dynamically unstable structure.

\subsection{Suitability Screening}
\subsubsection{HSE Band Gaps and Band Edges}

%%%%%%%%%%%%%%%%%%%%%%%%%%%%%
% BAND GAP & EDGE FIGURE 
%%%%%%%%%%%%%%%%%%%%%%%%%%%%%
 \begin{figure}[htb!]
 \begin{centering}
 \includegraphics[width=14cm]{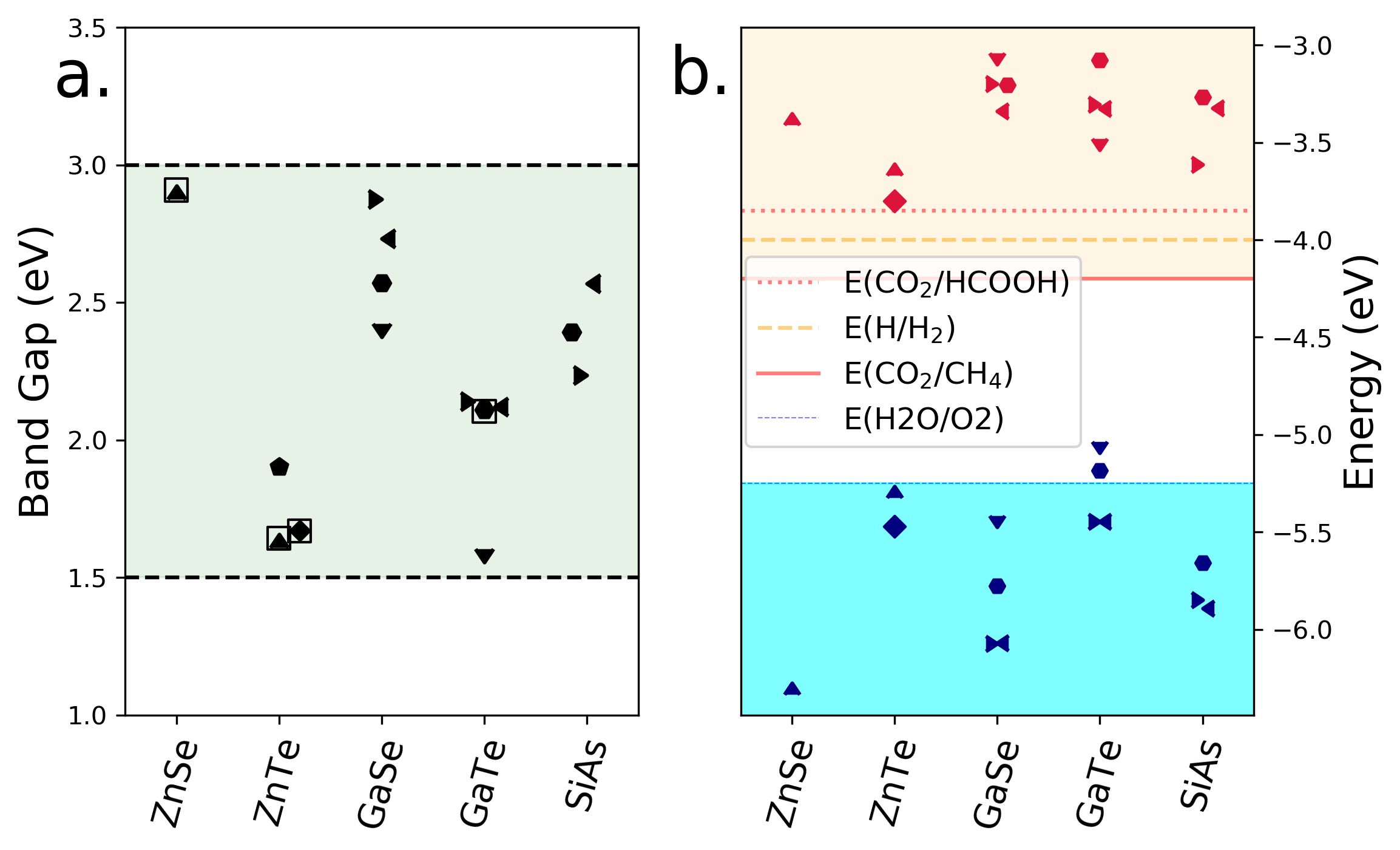}
 \caption{a. The HSE06\cite{Heyd2003,Krukau2006} band gaps tend to lie within the visible light spectrum. Boxed points indicate direct band gaps. The y-axis denotes band gap values. b. The band edges fulfill the requisite energy for \coo reduction\label{fig:electronics}. Blue (red) denotes valence (conduction) band maxima (minima). 
On both subplots, the x-axis groups band gaps and edges by material. 
Reaction energies on right from Figure 3 in Singh \latin{et al}.\cite{Singh2019}. 
}
\end{centering}
\end{figure}

%% WHY LOOK AT BAND GAPS?
We computed the electronic structure properties for all structures which passed the stability screening, in order to determine band gaps and edges, which are relevant to catalytic application. The HSE06 hybrid exchange-correlation functional\cite{Heyd2003,Krukau2006} is known to exhibit an improved treatment of semiconductor bandgaps: in Figure \ref{fig:electronics}, we show that the band gaps of the 2D materials computed using HSE06 lie mostly within the visible light spectrum, and the band edges are appropriate for \coo reduction. We found the GaS + T-Type structure of GaSe was metallic. 

To the best of our knowledge, the electronic structure of most of the 2D structural prototypes in this study have not been explored computationally before. Tong \latin{et al.}\cite{Tong2014} using PBE found ZnTe in the tetragonal form to have a band gap of 0.88 eV, lower than our HSE06 gap of about 1.6 eV. This is to be expected, as the PBE functional is known to underestimate band gaps. We reproduce Bai's monolayer SiAs indirect Y-$\Gamma$ HSE gap of 2.39 eV,\cite{Bai2018} though Cheng \cite{Cheng2018} reports a direct gap of 2.353 eV. Li \emph{et al.}\cite{Li2015e} found a band gap of 3.4 eV for the same ZnSe structure using HSE06, about 500 meV higher than what we find.

%%%%%%%%%%%%%%%%%%%%%%%%%%%%%%%%%%%%%%%%%%%
%%%%%%%%%%%% EXCITONIC PARAGRAPH
%%%%%%%%%%%%%%%%%%%%%%%%%%%%%%%%%%%%%%%%%%%%
\subsubsection{Quasiparticle gaps and Exciton binding energies}

We calculated the quasiparticle energies  within the GW approximation for the self energy which captures the many-body effects in the electronic structure \cite{hedin70,hybertsen1986electron} of materials. To calculate the optical response of materials, we solve the Bethe-Salpeter equation (BSE), \cite{onida2002electronic,PhysRevB.62.4927,rohlfing1998electron} where we explicitly take the electron-hole interaction into account. The solution to the BSE can be used to compute the imaginary part of the dielectric function, which enables us to study the excitonic effects in the absorption spectrum (Figures describing the BSE absorption spectrum can be found in the supplemental information). Low exciton binding energy is desirable for photovoltaic applications, as it enables easier electron-hole separation resulting in a higher device efficiency. However, in two-dimensional materials such as TMDCs, oftentimes large exciton energies ($\sim$ 1 eV) \cite{ramasubramaniam2012large} have been observed, arising from large effective charge carrier mass, strong Coulomb interactions, and weak dielectric screening \cite{Xiao2017} among other factors.

In Table \ref{table:excitonics} we show the quasiparticle gap (QPG), optical gap (OPG), and exciton binding energy (EBE) obtained from GW-BSE calculations. The QPG values we report here are the minimum direct bandgap for these materials, as we have not included any optical transitions with finite momentum transfer in our BSE calculations. The optical gap values reported in the table are the lowest excitation energies obtained by diagonalizing the BSE Hamiltonian. The EBE is then computed as the difference between the QPG and OPG. Our calculation shows for most of the materials the EBEs are large, similar to TMDCs. However, for a few of them we find low EBEs, such as 0.43 eV for ZnTe in the compressed SiAs-structure, 0.55 eV for SiAs in the native SiAs-structure, and 0.58 eV for GaTe in the GaS-structure.

\begin{table}
\begin{tabular}{|m{3cm}|m{4cm}|m{2cm}|m{2cm}|m{2cm}|}
\hline
\rowcolor[HTML]{C0C0C0}
Material & Structure Type & QPG (eV) & OPG (eV) & EBE (eV)\\
\hline
GaSe & GaS  &  3.89 &  3.22 &  0.67\\
\hline
\rowcolor[HTML]{EFEFEF}
GaSe & GaSe  &  4.05 &  3.38 &  0.68\\
\hline
GaSe & InSe  &  4.20 &  3.24 &  0.96\\
\hline
\rowcolor[HTML]{EFEFEF}
GaSe & Native SiAs  &  3.91 &  3.19 &  0.72\\
\hline
GaTe & GaS  &  3.37 &  2.79 &  0.58\\
\hline
\rowcolor[HTML]{EFEFEF}
GaTe & GaSe  &  3.58 &  2.90 &  0.68\\
\hline
GaTe & InSe  &  2.92 &  2.19 &  0.72\\
\hline
\rowcolor[HTML]{EFEFEF}
SiAs & GaS  &  3.67 &  3.06 &  0.61\\
\hline
SiAs & GaSe  &  3.57 &  2.92 &  0.66\\
\hline
\rowcolor[HTML]{EFEFEF}
SiAs & Native SiAs  &  1.88 &  1.33 &  0.55\\
\hline
ZnSe & Hexagonal  &  4.32 &  3.43 &  0.90\\
\hline
\rowcolor[HTML]{EFEFEF}
ZnTe & Hexagonal  &  3.06 &  2.32 &  0.74\\
\hline
ZnTe & Tetragonal  &  3.06 &  2.20 &  0.86\\
\hline
\rowcolor[HTML]{EFEFEF}
ZnTe & Compressed SiAs  &  2.97 &  2.54 &  0.43\\
\hline
\end{tabular}
\caption{Quasiparticle gap (QPG), Optical gap (OPG) and Exciton Binding Energy (EBE) computed for all structures. Full computational details are included in the supplementary information.}
\label{table:excitonics}
\end{table}

%% ELECTRONICS RESULTS DISCUSSION

The GW-BSE optical band gaps predicted for hexagonal ZnTe are in the visible light range (2.32 eV); those for SiAs in its native structure are slightly lower (1.33 eV) and for hexagonal ZnSe, slightly higher (3.43 eV).
Notably, transition metal chalcogenides can experience dramatic changes in opto-electronic behavior in the single-layer limit\cite{Mak2010}. For example, the photocurrent density of monolayer ZnSe increases dramatically\cite{Sun2012a} compared to the bulk. However, among the materials studied here, all but one of the 2D structures are predicted to be semiconductors, similar to their bulk counterparts. The HSE band gaps tended to increase for for SiAs, ZnSe, and GaSe, and were slightly lower for ZnTe\cite{Singh2019}.  Additionally, we find that ZnSe and ZnTe are, like their bulk parent structures, direct band-gap visible light semiconductors\cite{Singh2019}, which suggests promising potential for efficient light absorption. Monolayer GaSe, as in its bulk form, exhibits a visible-light indirect band gap\cite{Singh2019}. In the bulk, GaSe and SiAs have been identified as a material of interest for photovoltaic applications\cite{AndresPenares2017,2019arXiv190306651C}. On the other hand, GaSe and GaTe were recently reported to exhibit a sharp decrease in photoluminesence during the transition from the many to few-layer limit \cite{Pozo-Zamudio2015}.

\subsubsection{Adsorption and Reactivity}
%%%%%%%%%%%%%%%%%%%%%%%%%%%%%
% FREE ENERGY FIGURE
%%%%%%%%%%%%%%%%%%%%%%%%%%%%%
 \begin{figure}[htb!]
 \begin{centering}
 \includegraphics[width=14cm]{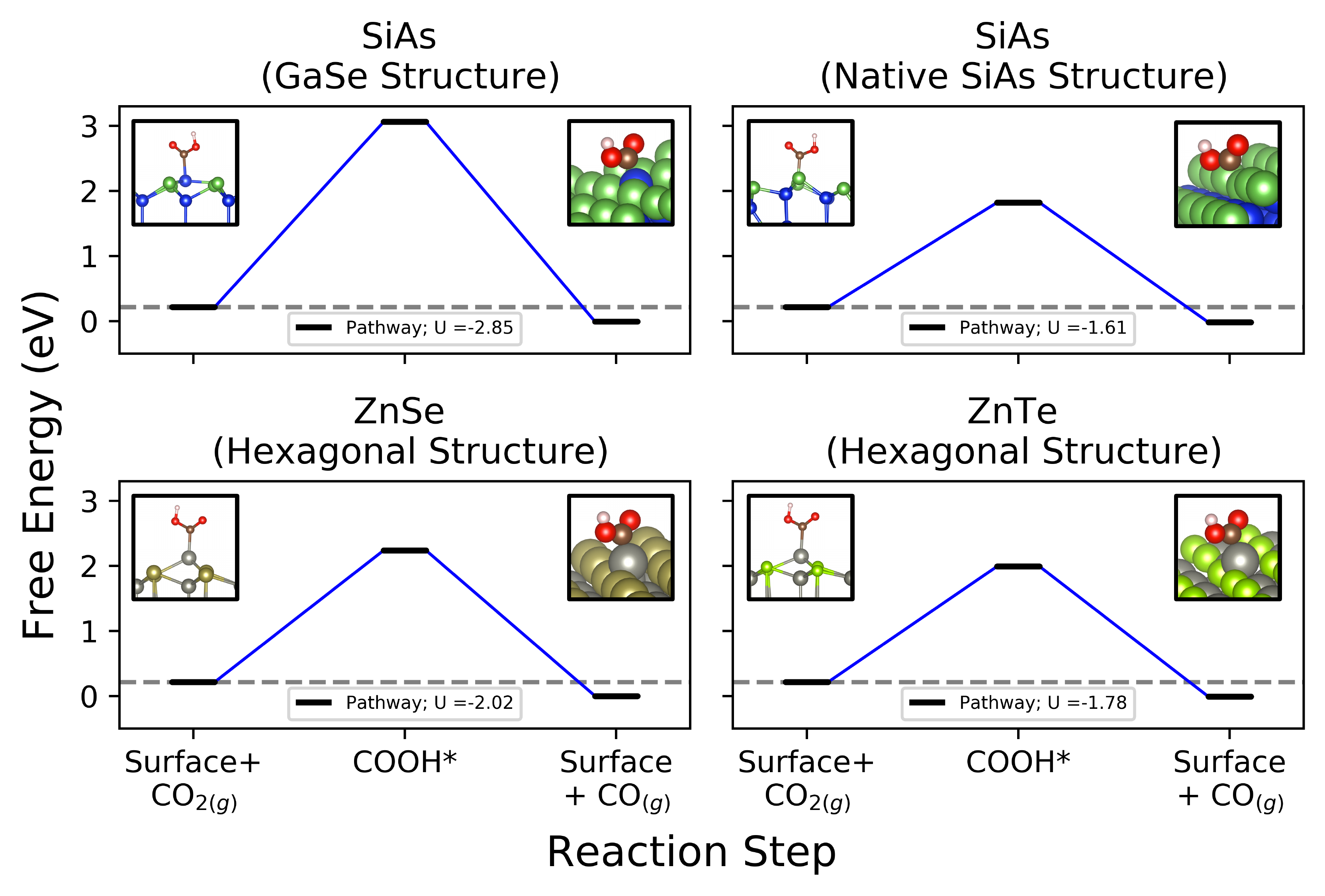}
\caption{ 
Free-energy diagrams for COOH bonding on SiAs, ZnTe, and ZnSe present relatively unreactive surfaces. Inset are images from the side (upper left) and at an angle (upper right) of COOH bonding on surfaces; the reconstruction is clearly visible when contrasting the side view with those from Figure \ref{fig:structures}.
Structure types are from Figure \ref{fig:structures}. SiAs in the native structure presents the lowest barrier.
For complete computational details, see supplemental information. 
\label{fig:free_energy}
}
\end{centering}
\end{figure}
% WHY LOOK AT ADSORPTION?
Finally, we studied adsorbate binding on the surfaces, which allows us to gauge the surface reactivity of the candidate photoelectrocatalyst  monolayers. \coo reduction can proceed along several complex reaction pathways \cite{Peterson2010,Kortlever2015,Schouten2011,Garza2018}. 
We examined the reaction pathway which predicts reactivity towards methane on copper \cite{Peterson2010}; though, in finding that CO does not bind to the surfaces, we focus only on the steps in equations (\ref{pathway1}) and (\ref{pathway2}) below, where * represents an arbitrary surface.

\begin{equation}
\text{CO}_{2(g)} + (H^+ + e^-) + * \rightarrow \text{*COOH} 
\label{pathway1}
\end{equation}

\begin{equation}
\text{*COOH} + (H^+ + e^-)  \rightarrow \text{CO}_{(g)} + \text{H}_2\text{O}_{(g)} + *
\label{pathway2}
\end{equation}

From adsorbate binding energies, we can compute the theoretical overpotential, which estimates the bias voltage that must be applied to the electrode in order for the reaction pathway to occur entirely downhill in free energy; this is equal to the greatest change in free energy.
Note that under photoelectrocatalyst operating conditions, energy which facilitates the reaction would be provided by the photoexcited electron, instead of the reaction being made spontaneous by an applied bias voltage. 
In this study we examine reactivity `in the dark' to gain insights on reaction selectivity and binding propensity; the full mechanisms of photoexcited electron charge transfer from the surface are beyond the scope of this study.
 The studied surfaces were the pristine basal planes of the candidate materials, with no defects nor edges. 
%, as we did not consider defects or surface edges.
Figure \ref{fig:free_energy} presents the resulting binding energies and the free energy reaction pathways for ZnTe, ZnSe, and SiAs. 
We focused on ZnTe, ZnSe, and SiAs for this phase of our study, because of the experimental evidence that monolayer GaTe and GaSe and absorb light poorly in the monolayer limit \cite{Pozo-Zamudio2015}.

% BINDING DISCUSSIOn
Examination of the bonding behavior of COOH and CO on the surfaces showed interesting results. For four structural prototypes (SiAs: GaSe and GaS, ZnSe: Hexagonal, and ZnTe: Hexagonal) we found that COOH binds via inducing a surface reconstruction of a Si or Zn atom, as the COOH radical `pulls' the cation out-of-plane. 
This behavior can be seen in the inset plots of  figure \ref{fig:free_energy}. We rationalize this for Zn as the Zn atom maintaining fourfold coordination by bonding with the adsorbate over the Se/Te atom below it; we also find this induces the Se/Te atom directly below the binding Zn to shift down and out-of-plane on the bottom.
For tetragonal ZnTe, we found that COOH did not bind to the surface, and the reconstruction seen in the hexagonal structure in which the Zn atom `pops' out-of-plane was not observed.
This suggests reactivity may be improved by the presence of vacancy or adatom defects which expose an undercoordinated Zn or Si atom out-of-plane.
Interestingly, for the native SiAs structure, the most favorable bonding environment is COOH adsorbed to the As atom highest out-of-plane, serving as a lower energy configuration than bonding to any Si atom. 

For every prototype we examined, we found no evidence of CO bonding to the surface; during relaxations of CO instantiated on various sites, CO uniformly shifted to far away from the surface. The `binding energies' of these configurations were approximately near zero. We examined the hydrocarbon-forming pathway from CO$_2$ to CHO \cite{Peterson2010}, but the lack of CO binding effectively terminates it at the second step. This suggests high selectivity for CO production, which could be used with a co-catalyst to facilitate other catalytic processes of interest.

% OVERPOTENTIAL DISCUSSION
While some 2D materials like TaS$_2$ and NbS$_2$ have been experimentally found to exhibit high basal plane reactivity \cite{Liu2017d}, the pristine structures we examined exhibit relatively unreactive overpotentials in the range of 1.61-2.02 eV, with SiAs in the GaSe structure around 2.85 eV. However, these surfaces may still hold promise for catalytic application, as deviation from the pristine structure may increase reactivity. The well known two-dimensional catalyst MoS$_2$ has an inert basal plane which can greatly increase in reactivity through defect engineering\cite{Li2015f,Ouyang2016}. Due to the lack of CO binding, we predict high selectivity towards CO. Weakly bound CO is critical to favorable CO production on Au \cite{Hansen2013,Feaster2017}, and the prediction of weak CO binding indicates that SiAs, ZnTe, and ZnSe belong to the class of materials characterized by the weak-binding leg of the volcano relationships derived in e.g. Kuhl \latin{et al}\cite{Kuhl2014a}.

%%%%%%%%%%%%%%%%%%%%%%%%%%%%%
%RECONSTRUCTION FIGURE
%%%%%%%%%%%%%%%%%%%%%%%%%%%%%
% \begin{figure}[htb!]
% \begin{centering}
% \includegraphics[width=16cm]{figures/reconstruction.pdf}
% \caption{
%The Zn atoms distort out-of-plane on ZnTe (left) and ZnSe (right) in order to bond with COOH. The chalcogen directly beneath %distorts out-of-plane in the opposite direction. Inset: Space-filling crystal structure at an angle for alternate perspective.
%\label{reconstruction}
%}
%\end{centering}
%\end{figure}

% DISCUSSION
\section{Conclusion}
In conclusion, our study examined the structural, thermodynamic, dynamic, electronic, and reactive properties of 2D forms of \coo reduction PECs. The primary contributions of this work are i) the uncommon chemistries of these materials in the \coo reduction literature, ii) our thorough exploration of possible monolayer phases, and iii) prediction of CO selectivity of SiAs, ZnSe, and ZnTe monolayers. 

 In native SiAs, hexagonal ZnTe, and hexagonal ZnSe, the comparatively lower overpotentials in tandem with ZnTe and ZnSe's direct band gaps present targets for further theoretical investigation, experimental verification, and property optimization. Possible means of engineering the catalytic activity of these structures includes tuning their reactivity and optical properties via defects, such as vacancies, adatoms, or dopants \cite{Ersan2016}. Heterostructuring could simultaneously allow for tuned opto-electronic properties or improved reactivity (as has been done for a tetragonal form of ZnSe with PbO) \cite{Zhou2018a}. In particular, the predicted high CO selectivity for SiAs, ZnSe, and ZnTe suggests that the presence of a co-catalyst could facilitate further reactions. Further excited-state studies could probe possible photoexcitation and reaction pathways. \cite{Kolesov2015} For ZnTe and ZnSe in particular, the projected density of states indicate that the high valence bands tend to exhibit anionic (Se, Te) character  and the low conduction bands have the character of the cation, Zn, similar to behavior seen in oxide systems\cite{Newhouse2017}. Closer examination of the interactions between individual adsorbates and these states will be the subject of further study.

In conclusion, novel 2D chemistries and structures for \coo photoreduction are suggested and explored computationally. The structures we studied in this paper preserve their bulk counterparts' stability and semiconductor properties. Further enhancement of surface reactivity remains a challenge, and further work is needed to understand how chemical changes, defect engineering and surface treatments can be used to influence and tune the performance. 

\section{Methods}

We performed all first-principles calculations in the \emph{Vienna Ab-Initio Simulation Package} with the Projector Augmented-Wave method \cite{Kresse1993,Kresse1996,Kresse1996a,Kresse1999}. Phonon calculations were assisted by Phonopy \cite{Togo2015a} with high cutoff energies at and above 700 eV, and supercells ranging in size from 5x5x1 to 8x8x1.  Automated workflows for relaxation, band gap estimation, and adsorption \cite{Montoya2017} were performed using Atomate \cite{Mathew2017}, using standard Materials Project parameters \cite{Jain2011,Jain2013} with small modifications specified in the SI. Structure matching, mapping, and general calculation IO operations were performed with pymatgen \cite{Ong2013} and fireWorks \cite{Jain2015}. Work function analysis was performed using pymatgen's Surface Analyzer package \cite{Sun2013,Tran2016}. More details on the calculation parameters, reference energies, and analysis are available in the SI.

For the energy calculation depicted in Figure \ref{fig:energetics}, we first relax structures with the PBE functional \cite{Perdew1996b}, then PBE+DFT-D3\cite{Grimme2010}, and again with SCAN+rVV10\cite{Peng2016}, a Meta-GGA functional with a van der Waals correction. We compare the energies with bulk structures using SCAN+rVV10 because it has shown good performance for estimating exfoliation energies of layered structures\cite{Tawfik2018}. Due to its improved prediction of band gaps \cite{Garza2016}, we relaxed the structures and computed the electronic structure using the HSE06 hybrid functional \cite{Heyd2003,Krukau2006}.
Exciton calculations were performed in VASP.

\section{Supporting Information}
The supplied supporting information contains full computational details for the thermodynamic, electronic, phonon, and adsorption calculations, an explanation of our analysis, full electronic and phonon band structures, and the frequency dependence of the imaginary $\epsilon_2$ component of the dielectric function.

\section{Acknowledgements}
Work on the photocatalyst screening was primarily funded by the Joint Center for Artificial Photosynthesis, a DOE Energy Innovation Hub, supported through the Office of Science of the U.S. Department of Energy under Award Number DE-SC0004993. Additional support was provided by the Materials Project (Grant No. KC23MP) through the DOE Office of Basic Energy Sciences, Materials Sciences, and Engineering Division, under Contract DE-AC02-05CH11231. 
Computations in this paper were run on the Odyssey cluster supported by the FAS Division of Science, Research Computing Group at Harvard University, on the National Energy Research Scientific Computing Center, a DOE Office of Science User Facility supported by the Office of Science of the U.S. Department of Energy under Contract No. DE-AC02-05CH11231, and the San Diego Supercomputer Center under the XSEDE Award No. TG-DMR150006. 
A.K.S. is supported by the National Science Foundation under Award No. DMR-1906030. S.B.T. is supported by the DOE Computational Science Graduate Fellowship under grant DE-FG02-97ER25308. 

S.B.T. would like to thank Wei Chen, Tara Boland, David Lim, and Jennifer Coulter for helpful discussions, as well as the Persson group for their hospitality during his visit.

\section{Author Contributions}
S.B.T., A.K.S., J.H.M., and K.A.P. jointly conceived the study. S.B.T. completed all thermodynamic, dynamic, electronic, and reaction calculations, and prepared the manuscript. A.K.S. and J.H.M. contributed guidance on calculations and analysis. T.B. contributed GW calculations and wrote the relevant sections of the manuscript. K.A.P. supervised the work. All authors provided feedback on the manuscript and discussed the results.

\section{Competing Interests}
The authors declare no competing interests.

%\begin{figure}
%\includegraphics{figures/TOC.pdf}
%\caption{For table of contents only.}
%\end{figure}

\bibliography{2DPC_Biblio.bib,extra.bib}

\end{document}